%% file: main-arxiv.tex
\documentclass[11pt]{article}

\usepackage[margin=1in]{geometry}

\usepackage{amsmath}
\usepackage{amssymb}
\usepackage{amsthm}
\usepackage{mathtools}
\usepackage{float}
\usepackage{graphicx}
\usepackage{booktabs}
\usepackage{microtype}
\usepackage{xcolor}
\usepackage{booktabs,tabularx,hyperref}
\usepackage[round,authoryear]{natbib}

\input{macros}

\title{Mutual Evaluation and Supervision without Peers}
\author{
  Zachary Robertson \\
  Department of Computer Science, Stanford University \\
  \texttt{zroberts@stanford.edu}
}
\date{}

\begin{document}

\maketitle

\begin{abstract}
\input{abstract}
\end{abstract}

\input{body}

\bibliographystyle{plainnat}
\bibliography{references}
\newpage
\input{sections/appendix}

\end{document}

%% file: macros.tex
\newtheorem{theorem}{Theorem}

\newtheorem{corollary}[theorem]{Corollary}

\theoremstyle{definition}

%% file: abstract.tex
% SOURCE: PLAN, line 8. Verbatim prose; LaTeX conversion only.
% CHECK: Reconcile the joint-report-law and timing language with the final core.

This article introduces mutual evaluation of a replicable task worker and a critic that incentivizes truthful reporting, both modeled as strategic agents. The critic chooses a finite-valued rule that induces an evaluation score on joint report laws. Their common payoff is analyzed through regret relative to the unrestricted critic envelope. The critic rule is distinct from the evaluation score.  This class enables a peer-free information elicitation mechanism using conditionally independent replications of a worker on the same task. This replication-loop mechanism implements a type-agreement payoff using same-task replications and new-task samples. In contrast to the peer-prediction and scoring-rule literature, implementations are shown that produce unbiased Pearson and Shannon information scores without requiring peers, a ground-truth reference, or likelihood-ratio estimation. A valid binary critic also can be represented by shared finite type annotations of worker returns. One runtime restriction is that the number of required replicas is random and can depend on the critic rule. Other timing effects, such as commitment and reoptimization, yield distinct incentives, connecting the framework to variational peer prediction. This mechanism class illustrates why strategic considerations matter for both critic and worker agents.

%% file: body.tex
% Manuscript order lives here, not in filenames.

\input{sections/introduction}

% Minimal common game language:
% workers, critic rule, outcome law, robustness, Nash.
\input{sections/framework}

% Restrict/represent the critics used by the replication mechanism.
\input{sections/annotations}

% Main construction: sampling interface, Pearson loop, KL loop,
% termination/integrability/unbiasedness.
\input{sections/replication}

% What the replication scores optimize:
% information envelopes, posterior regret, sufficiency,
% critic garbling, and truthful equilibrium.
\input{sections/information-and-incentives}

% General abstract envelope/regret facts.
% These are useful beyond replication, especially for the timing comparison,
% but are not prerequisites for understanding the mechanism itself.
\input{sections/value-and-regret}

% CA/VPP timing, commitment vs. reoptimization,
% worked Boolean example, and eventual peer-prediction separation lemma.
\input{sections/peer-prediction}

% Runtime/sample-cost caveats and unresolved design choices.
\input{sections/efficiency}

% Activate once the related-work prose is written.
% \input{sections/related-work}

\input{sections/conclusion}

% Functionally appendix-like, but intentionally retained in the paper body.
\input{sections/reproducibility}

%% file: sections/introduction.tex
\section{Introduction}
\label{sec:introduction}

Peer prediction and proper scoring rules are mechanisms that incentivize workers to
truthfully report their beliefs for completing evaluation tasks. Peer prediction
determines rewards from the reports of other workers rather than ground truth
\citep{miller2005peer}. Proper scoring rules score reported beliefs against (future)
observed outcomes \citep{brier1950verification}. Are there incentivizing mechanisms
that need neither peer workers nor (future) observed outcomes?

This article provides a positive answer. I formalize
(\href{https://github.com/zrobertson466920/mutual-evaluation/tree/main}{see Lean 4 repository})
mutual evaluation of a replicable task worker and a critic that compares returns,
both modeled as strategic agents seeking a common payoff (evaluation score). This model assumes access to a worker
whose behavior can be independently replicated on the same task. For example, a stochastic computational worker
can be run multiple times independently under the same task and specification. I then introduce a replication
loop implementation, based on waiting for critic matches, that returns an unbiased
estimate (correct in expectation) of the worker-critic payoff.

To be brief, replication \(Y_1'\) of \(Y_1\) on the same task \(X\) replaces the peer
\(Y_2\) as a task proxy. However, by the Data-Processing Inequality (DPI) mutual information between peers or
replicas would only lower-bound true task information \citep{schoenebeck2021variational}. By comparing
repeated replication attempts on the same task and freshly sampled tasks, the
mechanism is able to score how much the worker preserves true task information.

\[
\underbrace{I(Y_1;Y_2)}_{\text{peer proxy}}
\ \mathrel{\underset{\mathrm{DPI}}{\le}}\
\underbrace{I(X;Y_{1})}_{\text{true task information}}.
\]

This means producing a score does not require a second peer worker, a ground-truth
reference, or an estimator for likelihood ratios. For Pearson and Shannon
information, simple first-replication waiting times give unbiased estimates. In
terms of incentives, consider a worker that truthfully reports and a critic that
defines replication agreement without losing distinctions with the task
information. This worker-critic pair jointly maximizes the score and forms a Nash equilibrium. The
trade-off is a random, potentially unbounded number of worker replications. Additionally,
replication is a distributional hypothesis: distinct systems may serve as
replicas when they induce the same fixed conditional annotation channel. Mutual
evaluation also makes the incentives of the critic explicit, as it decides which worker returns count as the same.

\paragraph{Related work.}
The replication-loop estimators use inverse-binomial and waiting-time identities related to prior work on unbiased information-theoretic estimation \citep{degroot1959unbiased, vanopheusden2020ibs} and, more recently, peer-prediction sample complexity \citep{aznag2026sample}. The contribution here is not the waiting-time identity itself, but its use in a different sampling and strategic setting: independent replications of one worker on the same task, together with fresh-task samples, implement Pearson and Shannon information retained about the task. In particular, the replicas are not treated as peers and replica--replica information is not the evaluation objective; see the \hyperlink{scope-remarks}{scope remarks in Section~4}. The resulting mechanism makes the critic's notion of agreement a strategic annotation choice and permits its information loss and incentives to be analyzed explicitly.

\hypertarget{motivating-example}{}
\paragraph{Motivating example.}
The distinction from replica--replica information can be seen with a simple example. Let \(X\) be a fair bit and let independent replicas satisfy

$$
Y=X\oplus N,\qquad Y'=X\oplus N',
$$

where \(\oplus\) is the xor mapping rule and \(N,N'\) are independent Bernoulli noise bits with parameter \(\varepsilon\). Since \(Y\) is fair and \(H(Y\mid X)=h(\varepsilon)\),

$$
I(X;Y) := H(Y)-H(Y\mid X)
= \log 2-h(\varepsilon),
$$

where \(h\) is the binary entropy function with natural logarithms \cite[p.~13]{cover1991elements}. Moreover,

$$
Y\oplus Y'=N\oplus N',
$$

so \(Y\) and \(Y'\) disagree with probability \(2\varepsilon(1-\varepsilon)\). Therefore

$$
I(Y;Y')
=\log 2-h\!\left(2\varepsilon(1-\varepsilon)\right),
$$

which is strictly smaller for \(0<\varepsilon<1/2\). Mutual information between replicas can therefore fail to equal task information.

The formalism introduced here abstracts the empirical type-based mutual evaluation games of \citet{robertson2025step} by separating a critic rule from the evaluation score it induces. This separation makes a global validity condition on the critic analytically useful: a valid binary critic is representable as equality of finite annotations, so its information loss can be characterized directly. The same rule--score separation also allows committed and reoptimized critics to be compared through a common value envelope.

%% file: sections/framework.tex
\section{Mutual evaluation}
\label{sec:framework}

A mutual evaluation game \(G\) has an evaluation type (task set) \(X\) and a
completion return type (return alphabet) \(R\). Both are finite here. Write
\(\Delta(R)\) for beliefs over returns. A channel or kernel affects an input to a
belief (probability distributions) over outputs. A task \(x\sim P\), drawn from
the task prior (distribution) \(P\), affects raw returns through two fixed
worker channels

$$
w_i:X\to\Delta(R), \qquad i\in\{1,2\}.
$$

The critic chooses a finite-valued rule that induces an evaluation score on joint report laws.
This generalizes the game-theoretic separation between critic rule and induced evaluation score
used in the empirical type-based mutual evaluation games of \citet{robertson2025step}. The rule is

$$
c:R\times R\to S,
$$

with \(S\subset\mathbb R\) finite and nonempty; \(u(c,\rho)\) is the common payoff.
Write
\[
\mathcal C:=\{c:R\times R\to S\}
\]
for the critic rule set.

Each worker chooses a reporting kernel

$$
\sigma_i:R\to\Delta(R):
$$

$$
y_i\sim w_i(x),\qquad r_i\sim\sigma_i(y_i).
$$

A truthful strategy yields the raw return unchanged. Write
\[
\operatorname{truth}_i(y):=\delta_y,
\qquad
\operatorname{truth}:=(\operatorname{truth}_1,\operatorname{truth}_2).
\]
A garbling post-processes the reported return through a further kernel. Conditional on \(x\), independently
sample twice from each reported channel. The strategy profile
\(\sigma=(\sigma_1,\sigma_2)\) thus generates an outcome law (distribution)
\(\rho=\operatorname{law}_G(\sigma)\) over
\[
(R\times R)\times(R\times R).
\]
Write \(\rho_\sigma:=\operatorname{law}_G(\sigma)\).

Given an outcome law \(\rho\), define the critic envelope or supremal game valuation
\[
V(\rho):=\sup_{c\in\mathcal C}u(c,\rho).
\]

The game is weakly robust if worker garbling cannot increase the supremal game valuation:
if \(\nu\) is obtained from \(\rho\) by further garbling a worker's reported
return, then
\[
V(\nu)\le V(\rho).
\]

Peer prediction would use both worker
channels. The conceptual move in this article is to ignore the second worker
channel and design the evaluation score as a strategic channel.

%% file: sections/annotations.tex
% SOURCE: BLOG, lines 37--60.
% Additional representation bound: PLAN, line 284.

\section{Critic annotations}
\label{sec:annotations}

Here \(X\) also denotes the random task. For a finite-valued variable \(Z\), write
\(p_Z(z)=\Pr(Z=z)\) and \(p_{X,Z}(x,z)=\Pr(X=x,Z=z)\). Define Pearson and Shannon
mutual information by
\[
\begin{aligned}
I_{\chi^2}(X;Z)
&:=\sum_{x,z}\frac{p_{X,Z}(x,z)^2}{P(x)p_Z(z)}-1,\\
I(X;Z)
&:=\sum_{x,z}p_{X,Z}(x,z)
\log\frac{p_{X,Z}(x,z)}{P(x)p_Z(z)}.
\end{aligned}
\]

Terms with zero denominator are omitted, and zero-mass logarithmic terms contribute
zero. Logarithms are natural. The Shannon score is also called the
Kullback--Leibler (KL) score.

Fix the critic score set to \(S=\{0,1\}\). A critic rule
\(c:R\times R\to\{0,1\}\) can be seen as a relation. This relation is \emph{valid}
when
\[
y\sim_c y'
\quad\Longleftrightarrow\quad
c(y,y')=1
\]
defines an equivalence relation on the entire return alphabet. Notice validity is a
global property, not defined by a single sample.

\begin{theorem}[Finite annotation representation]
\label{thm:annotations}
Suppose \(R\) is finite. A critic is valid if and only if there are a finite
annotation alphabet \(B\) and a deterministic map \(g:R\to B\) such that
\[
c(y,y')=c_g(y,y')
:=\mathbf 1\{g(y)=g(y')\}
\qquad\text{for every }y,y'\in R.
\]
\end{theorem}

% SOURCE: PLAN, line 284.
The representation can be chosen with \(|B|\le |R|\).

% EDIT E6: "is way" -> "is a way".
A valid critic is a way to partition the worker's return alphabet. It decides which
returns count as the same type.

%% file: sections/replication.tex
% SOURCE: BLOG, lines 64--139.
% Score bookkeeping moved from BLOG, lines 157--162.
% Zero extension on nontermination: PLAN, line 367.
% CHECK: Formalization claims are retained, not independently audited.

\section{Replication loop mechanisms}
\label{sec:replication}

\subsection{Sampling interface}

The procedure introduced here only requires values of \(c(y,y')\) and a validity
check to run. This allows different annotation maps to induce the same
procedure. For both evaluation scores below, an invalid rule receives the constant
score zero and the procedure does not run. Intuitively, validity is a
\emph{property} not an assumption. If it is false the score is set to zero.

Replication should be interpreted as a distributional hypothesis rather than an
identity requirement: distinct systems may serve as replicas when their calls
are conditionally independent and induce the same fixed conditional annotation channel.
This preserves the replication-loop distribution, although treating the
systems as separately strategic agents would define a different game.

The critic here amounts to a finite annotation \(A=g(Y)\) of the worker's report
\(Y\), where \(g:R\to B\) maps returns to a finite label set \(B\). For the mechanism considered in this section, the two scores
are \(I_{\chi^2}(X;A)\) and \(I(X;A)\): Pearson and Shannon information retained
about the task. Worker garbling changes the reported information; critic garbling
coarsens the annotation by merging types.

Fix the single replicated worker's reported channel \(k=k_\sigma\), where
\(\sigma:R\to\Delta(R)\) now denotes this worker's reporting kernel applied after
\(w_1\). We will define two replication sequences using random variables based on a
fixed component. First sample the fixed task \(x\sim P\). Given \(x\), sample a
return \(Y\sim k(x)\) and use this as the fixed return, or anchor. The first
sequence consists of alternative return replications from the fixed task. These are
drawn according to
\[
\widetilde Y_n\sim k(x), \qquad n\ge1
\]

Conditional on \(x\), the law of the sequence
\((\widetilde Y_n)_{n\ge1}\) is independent and identically distributed (iid). The
second sequence consists of null return replications from freshly drawn tasks
sampled
\[
x_n\sim P,\quad Z_n\sim k(x_n), \qquad n\ge1.
\]

Conditional on the fixed task \(x\), the fixed return and replication sequences are
independent.

Given a valid critic define the alternative and null type-replication variables
(first-match counts, or hitting times) from the comparisons:
\[
\begin{aligned}
\tau_{\mathrm{alternative}}
&:=\inf\{n\ge1:c(Y,\widetilde Y_n)=1\},\\
\tau_{\mathrm{null}}
&:=\inf\{n\ge1:c(Y,Z_n)=1\},
\end{aligned}
\]
with \(\inf\varnothing=\infty\). The experiments below are based on inverse-binomial
sampling, but do not use a peer worker or ground truth \citep{degroot1959unbiased, vanopheusden2020ibs}.

\subsection{The Pearson collision payment}

Request one same-task return \(\widetilde Y_1\). If
\[
c(Y,\widetilde Y_1)=0,
\]
stop and pay \(-1\). Otherwise continue producing null replications and pay
\(\tau_{\mathrm{null}}-1\). Thus
\[
W_{\chi^2}
=
\begin{cases}
-1, & c(Y,\widetilde Y_1)=0,\\
\tau_{\mathrm{null}}-1, & c(Y,\widetilde Y_1)=1.
\end{cases}
\]

The null search is not run after a mismatch.

\subsection{The KL two-clock payment}

Run both replications (the two clocks). With harmonic numbers
\[
H_0:=0,\qquad H_m:=\sum_{j=1}^{m}\frac1j,
\]
pay
\[
W_{\mathrm{KL}}
:=
H_{\tau_{\mathrm{null}}-1}
-
H_{\tau_{\mathrm{alternative}}-1}.
\]

\hypertarget{scope-remarks}{}
\paragraph{Remarks.} The mechanisms presented here are not based on peer agreement. The terms $I(Y;\widetilde Y_n)$ are \textbf{not} generally equal to retained task information; recall \hyperlink{motivating-example}{the motivating example from the introduction}. Additionally, since the critic is strategic, it is assumed that the reporting channel and critic remain fixed throughout the experiment.

\subsection{Expected evaluation scores}

% SOURCE: BLOG, lines 157--162.
\paragraph{Channel-level evaluation scores.}
The parameter of the replication loop score is the channel instance \((P,k)\). In the replication game, a single worker chooses \(\sigma\) and the critic
chooses \(c\), with common expected payoff \(u_\bullet(c;P,k_\sigma)\). The
KL experiment need not be determined by the four-return outcome law used by the
core game.

Write \(\bullet\) for either \(\chi^2\) or \(\mathrm{KL}\), and \(\mathbb E\) for
expectation. This separates the critic rule, random payment, and evaluation score as
distinct constructs:
\[
c
\quad\longmapsto\quad
W_\bullet(c;P,k)
\quad\longmapsto\quad
u_\bullet(c;P,k)=\mathbb E[W_\bullet(c;P,k)].
\]

The replication loop mechanisms both terminate and equal their corresponding mutual
information functional in expectation. Specifically, the replication procedure
evaluates the information \emph{retained} by the equivalence classes of the critic.

\begin{theorem}[Termination, integrability, and unbiasedness]
\label{thm:replication}
For finite \(X,R\), every fixed channel \(k\), and every valid critic \(c\), each
invoked replication sequence terminates almost surely (with probability one) and
both payments are integrable (\(\mathbb E[|W_\bullet|]<\infty\)). If \(c=c_g\) and
\(A=g(Y)\), set \(A'=g(\widetilde Y_1)\) and \(p_A(a)=\Pr(A=a)\). Then
\[
\begin{aligned}
u_{\chi^2}(c;P,k)
&=
\sum_{a:p_A(a)>0}
\frac{\Pr(A=a,A'=a)}{p_A(a)}
-1
=
I_{\chi^2}(X;A),\\
u_{\mathrm{KL}}(c;P,k)
&=
I(X;A).
\end{aligned}
\]
\end{theorem}

Here \(A\) and \(A'\) are annotations of alternative replications, not annotations
of null replications. The proof is formalized, but the key calculation is simple.
Write

$$
q_x(a):=\Pr(A=a\mid X=x),\qquad p(a):=\Pr(A=a).
$$

Conditional on the fixed event \(X=x,A=a\), the alternative and null hitting
times satisfy

$$
\tau_{\mathrm{alternative}}\sim \operatorname{Geom}(q_x(a)),
\qquad
\tau_{\mathrm{null}}\sim \operatorname{Geom}(p(a)).
$$

Therefore

$$
\mathbb E[W_{\mathrm{KL}}\mid X=x,A=a]
=
-\log p(a)+\log q_x(a)
=
\log\frac{q_x(a)}{p(a)},
$$

using

$$
\tau\sim\operatorname{Geom}(p)
\quad\Longrightarrow\quad
\mathbb E[H_{\tau-1}]=-\log p.
$$

Similarly, conditional independence of the alternative replication and null search gives

$$
\mathbb E[W_{\chi^2}\mid X=x,A=a]
=
\frac{q_x(a)}{p(a)}-1.
$$

Averaging over \((X,A)\) yields \(I(X;A)\) and \(I_{\chi^2}(X;A)\), respectively.
This inverse-binomial identity is the basic mechanism by which replication times
become unbiased estimates of information \citep{vanopheusden2020ibs}.

%% file: sections/information-and-incentives.tex
% SOURCE: BLOG, lines 143, 165--169, 190--227.

\section{Information, sufficiency, and incentives}
\label{sec:information-incentives}

The replication loop mechanism evaluates the information \emph{retained} by the
equivalence classes of the critic. Sufficiency means preserving task beliefs when
replacing a report by its annotation. The value envelope and regret associated with
the mutual evaluation game will help clarify the incentives.

The envelope and critic regret are then the generic score quantities
\[
V_\bullet(P,k):=\sup_c u_\bullet(c;P,k),
\qquad
r_\bullet(c;P,k):=V_\bullet(P,k)-u_\bullet(c;P,k).
\]

\paragraph{Envelope and posterior regret.}
On the support of reports and annotations, define the conditional distributions of
the task given the report and its annotation.
\[
\pi_y(x):=\Pr(X=x\mid Y=y),
\qquad
\bar\pi_a(x):=\Pr(X=x\mid A=a).
\]

Since \(A=g(Y)\) we have an alternative representation in terms of conditional
mutual information.
\[
I(X;Y\mid A)
:=
\sum_{y:p_Y(y)>0}
p_Y(y)
\sum_x
\pi_y(x)
\log
\frac{\pi_y(x)}{\bar\pi_{g(y)}(x)}.
\]

Note zero-mass terms contribute zero. There are simple expressions for the game valuation and critic regret.

\begin{theorem}[Envelopes, regret, and sufficiency]
\label{thm:information-regret}
Maximizing over the full binary critic space gives
\[
V_{\chi^2}(P,k)=I_{\chi^2}(X;Y),
\qquad
V_{\mathrm{KL}}(P,k)=I(X;Y).
\]

Both envelopes are attained by the literal-agreement critic
\[
c_{\mathrm{id}}(y,y')
:=
\mathbf 1\{y=y'\}.
\]

For every valid critic \(c=c_g\), with \(A=g(Y)\),
\[
\begin{aligned}
r_{\chi^2}(c;P,k)
&=
I_{\chi^2}(X;Y)-I_{\chi^2}(X;A)\\
&=
\sum_{y:p_Y(y)>0}
p_Y(y)
\sum_{x:P(x)>0}
\frac{
\bigl(\pi_y(x)-\bar\pi_{g(y)}(x)\bigr)^2
}{
P(x)
},\\[1ex]
r_{\mathrm{KL}}(c;P,k)
&=
I(X;Y)-I(X;A)\\
&=
I(X;Y\mid A).
\end{aligned}
\]
\end{theorem}

One useful corollary is that for either score, a valid critic has zero regret
exactly when
\[
\pi_y=\bar\pi_{g(y)}
\qquad
\text{for every }y\text{ with }p_Y(y)>0.
\]

Critic regret measures what the annotation discards. In the KL case, because
\(A=g(Y)\), the mutual-information chain rule gives the regret identity  \cite[Theorem~2.5.2, p.~22]{cover1991elements}. This 
is the task information still present in \(Y\) after observing
its annotation \(A\). This leaves an optimal critic free to
merge reports that leave beliefs unchanged over tasks. This separates
type-annotation from information loss. Literal agreement always attains the
envelope. However, this need not be the only optimal critic. This is useful since
the runtime may depend on the complexity of the type-annotations.

\paragraph{Examples.}
Let \(X\) and \(U\) be independent fair bits and let the truthful return be
\[
Y=(X,U).
\]

Literal agreement retains the full report. The nuisance-removing annotation
\[
g(x,u)=x
\]
strictly compresses the return alphabet but retains all task information. Both
critics therefore attain Pearson information \(1\) and Shannon information
\(\log2\), and both have zero regret.

A constant annotation retains no task information, giving regrets \(1\) and
\(\log2\), respectively. Any bijective annotation merely relabels reports and again
has zero regret.

These examples illustrate that the following three notions are distinct: the size of
the annotation alphabet, the amount of task information retained by the critic, and
the sampling cost of implementing its evaluation score.

\subsection{Garbling and truthful equilibrium}

Type-annotation gives a second garbling operation. If \(c=c_g\) is valid and
\(h:B\to C\) maps annotations to another label set \(C\), then
\[
c_{h\circ g}(y,y')
=
\mathbf 1\{h(g(y))=h(g(y'))\}
\]
is another valid critic obtained by garbling the annotation classes. For either
score,
\[
u_\bullet(c_{h\circ g};P,k)
\le
u_\bullet(c_g;P,k).
\]

Here \(w_1\) is the raw channel of the single replicated worker. For every reporting
kernel and every critic \(d\),
\[
u_\bullet(d;P,k_\sigma)
\le
V_\bullet(P,k_\sigma)
\le
V_\bullet(P,w_1).
\]

% SOURCE: BLOG, line 575.
% This version states the corollary without repeating the Nash definition.
\begin{corollary}[Truthful equilibrium under replication]
\label{cor:truthful-replication}
Truth and any critic optimal at truth jointly maximize the common evaluation score
and form a Nash equilibrium of the corresponding replication game. Worker deviations
choose one reporting kernel before the replication experiment begins.
\end{corollary}

%% file: sections/value-and-regret.tex
% SOURCE: PLAN, lines 120--185.
% Planning statements are retained, not independently certified.

\section{Value envelopes and critic regret}
\label{sec:value-regret}

When the evaluation or return types \(X\) or \(R\) appear in this section they refer
to the mutual evaluation game \(G\). The term \((X,R)\) is referred to as the
evaluation task type. Define critic regret as the difference between the supremal
game valuation and critic evaluation score.
\[
r(c,\rho):=V(\rho)-u(c,\rho).
\]

\phantomsection
\label{claim:value-decomposition}

By definition,
\[
u(c,\rho)=V(\rho)-r(c,\rho).
\]
Since the critic rule set is finite and nonempty, the supremum is attained.
Consequently, regret is nonnegative, vanishes exactly at an optimal critic,
and some critic has zero regret at every outcome law. These facts do not
require weak robustness.

\phantomsection
\label{claim:committed-loss}

If a critic $c$ is optimal at $\rho$, then its payoff loss at a new law
$\nu$ decomposes as
\[
u(c,\rho)-u(c,\nu)
=
V(\rho)-V(\nu)+r(c,\nu).
\]
Thus a committed critic can lose payoff both because the optimized value
decreases and because it is no longer optimal at the new law.
Reoptimizing the critic removes the latter term.

% EDIT E3: "guarnatees" -> "guarantees".
So if \(\rho\) is generated and \(\nu\) comes from admissible worker garbling, the
robustness property guarantees the value loss is nonnegative. This shows a
value-preserving garbling can lower the committed payoff through the new-law
regret. Reoptimizing the critic removes the regret term while holding the critic
fixed need not.

%% file: sections/peer-prediction.tex
% SOURCE: PLAN, lines 191--260.
% The separation lemma/counterexample are not yet supplied.

\section{Comparison with peer prediction}
\label{sec:peer-prediction}

\subsection{Critic timing: commitment and reoptimization}
\label{sec:critic-timing}

We compare an idealized fixed-score correlated-agreement (CA) implementation with an actual-law-reoptimized variational peer-prediction (VPP) objective \citep{shnayder2016informed, schoenebeck2021variational}. This is intended to illustrate how in the mutual evaluation game they have the same evaluation task type and envelope, but different critic regrets. The comparison assumes perfect access to the outcome law to illustrate timing effects in a simple setting; this is a return to the two-worker outcome-law game of
Section~\ref{sec:framework}. The validity criteria used for the replication
scores do not apply to an arbitrary payoff \(u(c,\rho)\).

In the \textbf{CA timing}, the critic rule is selected assuming truthful reporting
and then held fixed. In the \textbf{VPP timing}, the critic rule is reoptimized based
on the actual outcome law. Write \(\rho_*:=\operatorname{law}_G(\operatorname{truth})\). For finite
\(R\), it is possible to choose a critic \(c_*\) that is optimal at \(\rho_*\).
\[
u(c_*,\rho_*)=V(\rho_*).
\]
If the actual outcome law shifts to \(\nu\) then the two timings generate evaluation
scores which can be written as:
\[
U_{\mathrm{CA}}(\nu):=u(c_*,\nu),
\qquad
U_{\mathrm{VPP}}(\nu):=V(\nu).
\]
They agree at truth. In general, the payoff difference is the critic regret of CA
\[
U_{\mathrm{VPP}}(\nu)-U_{\mathrm{CA}}(\nu)=r(c_*,\nu).
\]
This is the committed critic's regret in the original game.

The committed-loss identity makes critic timing relevant. Suppose the critic starts
as optimal. Weak robustness says if the primary worker garbles their completion belief
the optimized value (weakly) decreases. If we have weak robustness for both workers the same holds for garblings by either worker.

\begin{theorem}[Equilibrium transfer]
\label{thm:equilibrium-transfer}
If
\[
V(\rho_\sigma)\le V(\rho_*)
\quad\text{for every reporting profile }\sigma,
\]
then
\[
u(d,\rho_\sigma)\le u(c_*,\rho_*)
\quad\text{for every critic }d\text{ and profile }\sigma.
\]
Thus \((\mathrm{truth},c_*)\) jointly maximizes common payoff and is Nash.
\end{theorem}

\paragraph{A simple Boolean objective.}
For a concrete comparison we take \(R=\{0,1\}\) and specialize to CA and VPP with Total Variation divergence where it is known an optimal critic exists with
\(S=\{0,1\}\). For any outcome law \(\rho\), define \(\operatorname{cross}(\rho)\) by sampling
\[
\bigl((r_1^1,r_1^2),(r_2^1,r_2^2)\bigr)\sim\rho
\]
and returning \((r_1^1,r_2^2)\). Now define
\[
\begin{aligned}
\mu_\rho&:=\operatorname{cross}(\rho),\\
\pi_\rho(a,b)&:=(\mu_\rho)_1(a)(\mu_\rho)_2(b),\\
T(\rho)&:=\frac12\sum_{a,b}|\mu_\rho(a,b)-\pi_\rho(a,b)|,\\
J(c,\rho)&:=\mathbb E_{\mu_\rho}[c]-\mathbb E_{\pi_\rho}[c].
\end{aligned}
\]
CA evaluates a committed critic in this objective, while VPP reoptimizes the same
objective. The VPP mechanism, because the optimal critic is an element of the set of
all binary critics, can be instantiated with:
\[
\max_{c:\{0,1\}^2\to\{0,1\}}J(c,\rho)=T(\rho).
\]
The conclusion is that the maximum of the shared objective \(J\) equals the total
variation distance \(T\).

\paragraph{A worked example.}
Fix any task prior \(P\) and two raw Boolean-report worker channels. Define their
shared game payoff by
\[
u(c,\rho):=1+\frac12J(c,\rho).
\]
The envelope and original-game regret are
\[
V(\rho)=1+\frac12T(\rho),
\qquad
r(c,\rho)=\frac12\bigl(T(\rho)-J(c,\rho)\bigr).
\]
The game is an instance of mutual evaluation.

Additionally, for every choice of prior and worker channels, this game satisfies the weak robustness condition for the primary worker and truthful maximality \(V(\rho_\sigma)\le V(\rho_*)\) for every \(\sigma\). Therefore, the previous equilibrium-transfer theorem applies to any critic optimal at truth. Sample a uniform Boolean task bit and let both raw workers return that bit. Let \(\rho_*\) be the truthful outcome law and \(\nu\) the outcome law when only the first worker
flips its report. Choose the agreement critic
\[
c_{=}(a,b):=\mathbf 1\{a=b\}.
\]
The score and valuation calculation yields
\[
\begin{array}{c|cc}
 & \rho_* & \nu\\ \hline
U_{\mathrm{CA}}\text{ (agreement held fixed)} & 5/4 & 3/4\\
U_{\mathrm{VPP}}\text{ (reoptimized)} & 5/4 & 5/4
\end{array}
\]
Agreement attains the shared envelope at truth, so it meets the CA selection
condition. The bit flip preserves the optimized value but not the committed
evaluation score: CA falls from \(5/4\) to \(3/4\), while reoptimized VPP remains at
\(5/4\). The committed critic's regret rises from zero to \(1/2\); an actual-law
optimizer has zero regret.

% PLANNED SUBSECTION -- activate once the statement is supplied.
% \subsection{Separation from peer prediction}
% \label{sec:peer-separation}
%
% TODO: Insert the intended "this is not peer prediction" lemma and
% counterexample. Reuse the example setup rather than duplicating it.
% No new theorem statement or counterexample has been invented here.

%% file: sections/efficiency.tex
\section{Efficiency and limitations}
\label{sec:efficiency}

The remarks in this section are not formalized. Both mechanisms require a
fixed memoryless (history-independent) worker-critic strategy profile and may
use a random, unbounded number of samples. Expected sample count and estimator
variance are distinct criteria for practical efficiency.

For a valid critic, let \(m=|\operatorname{supp}(A)|\), and let \(N_\bullet\)
count the anchor report and all worker replications. KL runs two geometric
clocks after the anchor. Averaging their conditional means gives

$$
\mathbb E[N_{\mathrm{KL}}]
=
1+\mathbb E[1/p(A)]+\mathbb E[1/q_X(A)]
\le 1+2m.
$$

Pearson instead uses one replication for its same-task gate and runs a null
clock only after a match, giving

$$
\mathbb E[N_{\chi^2}]
=
2+\mathbb E\!\left[\frac{q_X(A)}{p(A)}\right]
=
3+I_{\chi^2}(X;A)
\le 2+m.
$$

Thus expected sample counts are finite on finite alphabets despite unbounded
realized waiting times.

Conditional on \((X,A)\), each KL harmonic term has variance at most
\(\pi^2/6\) \citep[Eq.~(15)]{vanopheusden2020ibs}. Conditional independence
therefore gives \(\operatorname{Var}(W_{\mathrm{KL}}\mid X,A)\le\pi^2/3\).
The conditional mean has finite variance on its finite support, so KL has
finite unconditional variance.

Pearson variance is also finite for each fixed instance but is not uniformly
bounded. Take \(A=X\sim\operatorname{Bernoulli}(\varepsilon)\). The same-task
gate always opens, and conditional on \(X=1\),

$$
W_{\chi^2}+1\sim\operatorname{Geom}(\varepsilon).
$$

Since \(\operatorname{Var}(\operatorname{Geom}(p))=(1-p)/p^2\), the law of
total variance gives

$$
\operatorname{Var}(W_{\chi^2})
\ge
\varepsilon\frac{1-\varepsilon}{\varepsilon^2}
=
\frac{1-\varepsilon}{\varepsilon}
\longrightarrow\infty,
$$

even though \(\mathbb E[N_{\chi^2}]=4\). Related sampling--variance tradeoffs
are discussed in \citet[Sections~4.3--4.4]{aznag2026sample}.

\paragraph{A timing caveat.}
After the anchor, advancing each unfinished clock once per parallel round gives
\(T_{\mathrm{KL}}=\max\{\tau_{\mathrm{null}},\tau_{\mathrm{alternative}}\}\),
not total sample count. On the same draws, Pearson's null-search wait,
excluding its initial same-task test, is
\[
T_{\chi^2}:=
\mathbf 1\{\tau_{\mathrm{alternative}}=1\}\tau_{\mathrm{null}}
\le T_{\mathrm{KL}}.
\]
Moreover, \(W_{\chi^2}=T_{\chi^2}-1\), whereas KL pays a harmonic difference:
using this gated raw wait as a payoff optimizes Pearson, not KL.

If the worker or critic adapts to elapsed time or replication history, the
fixed geometric-clock calculation need not apply. The memoryless profile is
therefore a substantive assumption, not merely an implementation convention.

%% file: sections/conclusion.tex
% SOURCE: BLOG, line 254. Only the heading changes.

\section{Conclusion}
\label{sec:conclusion}

The mutual evaluation model provides a common formalization for the peer-prediction
and replication settings. Peer prediction uses both worker channels. The conceptual
move here is to replace the second worker by independent replications of the first,
while treating the critic's notion of agreement as a strategic choice.

%% file: sections/reproducibility.tex
% SOURCE: BLOG, line 271. Original prose retained.
%
% CHECK: Reconcile the formalization claim with the actual theorem ledger.
% CHECK: Review "in this blog" and the authorship/process description for release.

\section{Formalization and reproducibility}
\label{app:formalization}

Unless otherwise noted, all mathematical declarations and statements in this article are formalized in Lean 4 and are available at \href{https://github.com/zrobertson466920/mutual-evaluation/tree/main}{this repository}\footnote{\url{https://github.com/zrobertson466920/mutual-evaluation/tree/main}}. Table \ref{tab:lean-correspondence} provides a correspondence between claims in the article and their formalized statements. Every theorem claim appearing in the article is derived from a set of manually reviewed formal declarations and associated annotations. The prose of the article was human authored and was written from these reviewed declarations and annotations. AI tools were used substantially to assist with the formalization of \emph{proofs}, but not with the formulation or review of theorem statements, and to help collate annotations during article preparation.

%% file: sections/appendix.tex
\appendix
\section{Formalization and Reproducibility}

% Stable links into the public Lean interface at the arXiv commit.
\newcommand{\leansource}[4]{%
  \href{https://github.com/zrobertson466920/mutual-evaluation/blob/9e294b264d89b1ef0bcc978244773d5cf932df39/MutualEvaluation/Public/#1.lean\#L#2-L#3}{#4}%
}

\begin{table}[H]
\centering
\small
\caption{Correspondence between manuscript statements and the public Lean
formalization. Links are pinned to commit
\texttt{9e294b264d89b1ef0bcc978244773d5cf932df39}.}
\label{tab:lean-correspondence}

\begin{tabularx}{\textwidth}{@{}p{0.19\textwidth}p{0.14\textwidth}X@{}}
\toprule
Manuscript\newline statement & Lean source & Principal declarations \\
\midrule

\hyperref[thm:annotations]{Theorem~\ref*{thm:annotations}}
&
\leansource{Replication}{70}{78}{\texttt{RP-02}}
&
\texttt{Binary.annotate\_valid};
\texttt{Binary.valid\_iff\_annotation}
\\[0.6em]

\hyperref[thm:replication]{Theorem~\ref*{thm:replication}}
&
\leansource{Replication}{477}{529}{\texttt{RP-09}}
&
\texttt{Replication.termination};
\texttt{integrable\_W\_chiSquared};
\texttt{integrable\_W\_KL};
\texttt{u\_chiSquared\_annotation};
\texttt{u\_KL\_annotation};
\texttt{u\_chiSquared\_collision\_ratio}
\\[0.6em]

\hyperref[thm:information-regret]{Theorem~\ref*{thm:information-regret}}
&
\leansource{Replication}{684}{724}{\texttt{RP-11}}
&
\texttt{V\_chiSquared\_eq};
\texttt{V\_KL\_eq};
\texttt{chiSquared\_literal\_optimal};
\texttt{kl\_literal\_optimal};
\texttt{r\_chiSquared\_annotation};
\texttt{r\_KL\_annotation};
\texttt{chiSquared\_annotation\_zero\_iff};
\texttt{kl\_annotation\_zero\_iff}
\\[0.6em]

\hyperref[cor:truthful-replication]{Corollary~\ref*{cor:truthful-replication}}
&
\leansource{Replication}{822}{832}{\texttt{RP-12}},
\leansource{Replication}{857}{877}{\texttt{RP-13}}
&
\texttt{chiSquared\_truthful\_global\_optimal};
\texttt{kl\_truthful\_global\_optimal};
\texttt{chiSquared\_truthful\_nash};
\texttt{kl\_truthful\_nash}
\\[0.6em]

\hyperref[claim:value-decomposition]{Value decomposition}
&
\leansource{Abstract}{8}{33}{\texttt{Abstract}}
&
\texttt{payoff\_eq\_value\_sub\_regret};
\texttt{regret\_nonneg};
\texttt{regret\_zero\_iff};
\texttt{regret\_attained}
\\[0.6em]

\hyperref[claim:committed-loss]{Committed-loss}
&
\leansource{Abstract}{92}{95}{\texttt{Abstract}}
&
\texttt{committed\_loss}
\\[0.6em]

\hyperref[thm:equilibrium-transfer]{Theorem~\ref*{thm:equilibrium-transfer}}
&
\leansource{CriticTiming}{59}{79}{\texttt{PROSE-05--06}}
&
\texttt{truthValueMaximal};
\texttt{truthful\_global\_optimal}
\\

\bottomrule
\end{tabularx}
\end{table}